\documentclass[final, twocolumn,10pt]{elsarticle}

\usepackage{amssymb}

\journal{Nuclear Instruments and Methods B}

\begin{document}

\begin{frontmatter}

\cortext[cor1]{peter.thirolf (-at-) physik.uni-muenchen.de}

\title{Towards a precise determination of the excitation energy of the Thorium nuclear isomer using a magnetic bottle spectrometer}

\author{Benedict Seiferle, Lars von der Wense, Ines Amersdorffer, Nicolas Arlt, Benjamin Kotulski \& Peter G. Thirolf\corref{cor1} }

\address{Ludwig-Maximilians-Universit\"at M\"unchen, Am Coulombwall 1, 85748 Garching, Germany.}

\begin{abstract}
$^{229}$Th is the only known nucleus with an excited state that offers the possibility for a direct laser excitation using existing laser technology.
Its excitation energy has been measured indirectly to be 7.8(5) eV ($\approx$160 nm).
The energy and lifetime of the isomeric state make it the presently only suitable candidate for a nuclear optical clock, the uncertainty of the excitation energy is, however, still too large to allow for a direct laser excitation in a Paul trap.
Therefore, a major goal during the past years has been an improved energy determination.
One possible approach is to measure the kinetic energy of electrons which are emitted in the internal conversion decay of the first isomeric state in $^{\mbox{\scriptsize 229}}$Th.
For this reason an electron spectrometer based on a magnetic bottle combined with electrical retarding fields has been built.
Its design, as well as first test measurements are presented, which reveal a relative energy resolution of 3 \%  and thus enable to measure the electrons' expected kinetic energy to better than 0.1 eV. 
This is sufficiently precise to specify a laser system able to drive the nuclear clock transition in $^{\mbox{\scriptsize 229}}$Th.
\end{abstract}


\end{frontmatter}


\section{Introduction}\label{Intro}

The isomeric nuclear first excited state in $^{229}$Th (called $^{229\mbox{\scriptsize m}}$Th) has the lowest excitation energy of all known nuclear states \cite{NNDC}.

The energy difference between the ground state ($J^\pi = 5/2^+$, with Nilsson numbers [633]) and the first isomeric state ($J^\pi = 3/2^+$, with Nilsson numbers [631]) has been measured indirectly to 7.8$\pm$0.5 eV \cite{Beck1, Beck2}, which corresponds to a wavelength of ~$\approx$160 nm and makes nuclear laser excitation possible.
The lifetime in $^{229}$Th ions  is expected to lie in the range of $10^3$ to $10^4$ s \cite{Ruchowska,Jeet} which leads to an exceptionally small natural linewidth of $\Delta E/E \approx 10^{-20}$.
Its unique properties stimulated the idea to use the isomeric transition as a nuclear frequency standard \cite{Peik1, Campbell1}.
As currently used optical atomic clocks, such a nuclear clock may find applications in a variety of fields, such as relativistic geodesy \cite{Mehlstaubler}, search for dark matter \cite{Derevianko} or time variation of fundamental constants \cite{Flambaum, Rellergert}.
A problem is, however, the large uncertainty in the knowledge of the excitation energy, which up to today impedes a direct laser excitation of the nucleus.
For this reason a major objective is a precise determination of the isomeric energy:\\
Worldwide several different approaches are pursued, either via indirect measurements of $\gamma$ rays emitted in the decay of higher lying states in $^{229}$Th \cite{Kazakov} or via direct optical excitation of the nucleus \cite{Jeet, Wense2, Stellmer2}.
Our goal is the spectroscopy of electrons which are emitted during the ground-state decay of $^{229\mbox{\scriptsize m}}$Th \cite{Seiferle2, Stellmer1}:

\begin{figure*}[ht]
\centering
\includegraphics[width=1\textwidth]{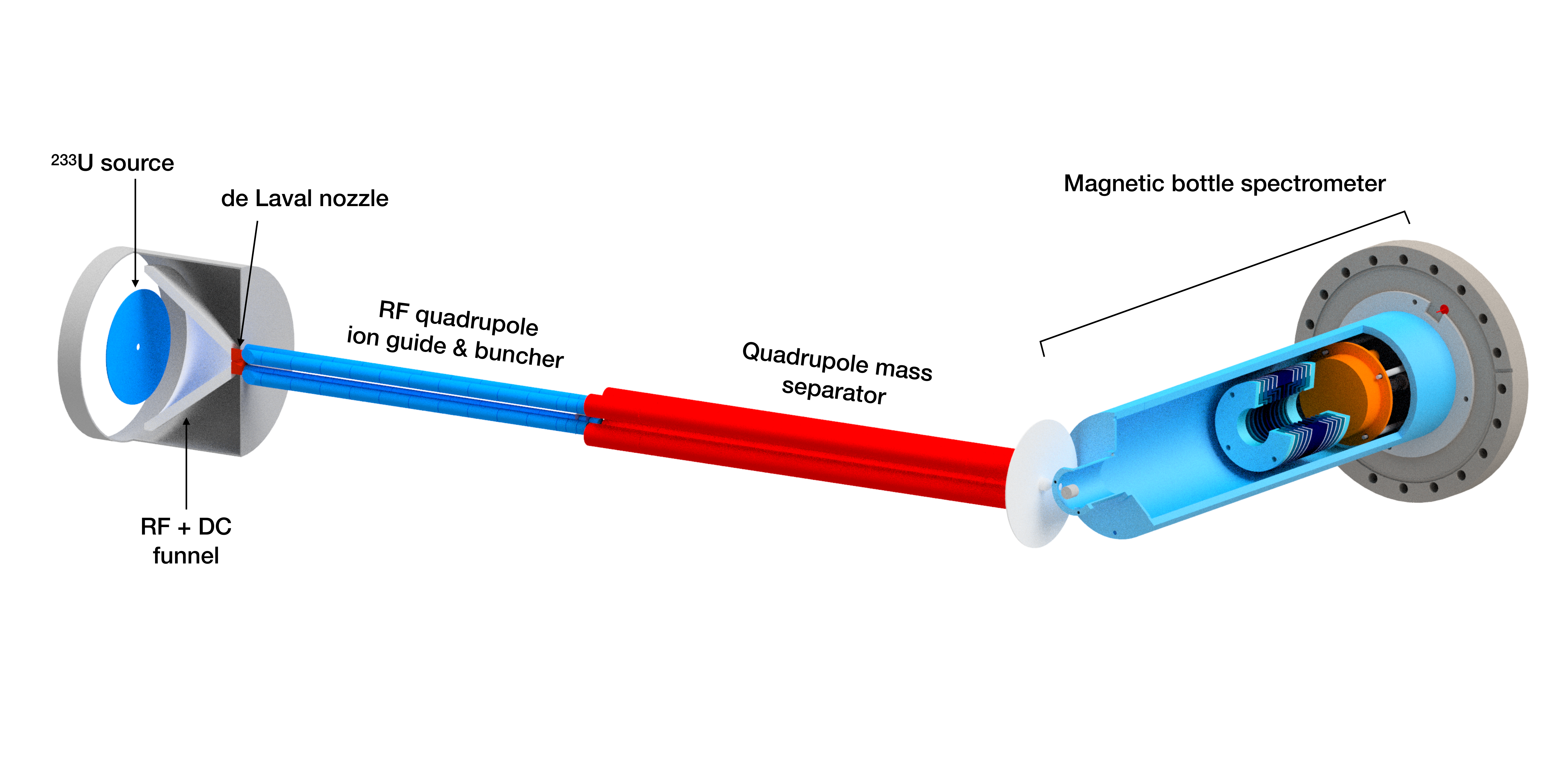}
\caption{The experimental setup which is used for a direct determination of the excitation energy of the isomeric state in $^{229}$Th. $^{229\mbox{\scriptsize m}}$Th ions are generated in the $\alpha$ decay of $^{233}$U. The source consists of a thin layer of $^{233}$U, such that decay products may leave the source material due to their recoil energy and form ions. These ions are thermalized in a buffer gas stopping cell, guided via an RF+DC funnel to a de-Laval nozzle and injected in an RF quadrupole ion guide, where ion bunches can be formed. Remaining daughter nuclei from the $^{233}$U decay chain, which are present in the ion beam (or bunch), are removed in a quadrupole mass separator.
The ions are guided to a magnetic bottle spectrometer, where they are neutralized and subsequently decay via internal conversion. \label{ExpSetup}}
\end{figure*}
In our experimental setup $^{229\mbox{\scriptsize m}}$Th is populated via a 2\% decay branch \cite{Thielking} in the $\alpha$ decay of $^{233}$U \cite{Wense}.
$^{229\mbox{\scriptsize m}}$Th recoil ions are thermalized in a buffer gas stopping cell and extracted into a subsequent segmented radiofrequency quadrupole (RFQ) ion guide.
If required, ion bunches can be formed in the RFQ \cite{Seiferle}.
A quadrupole mass separator allows to remove accompanying daughter nuclei from the $^{233}$U decay chain and select a specific (up to 3+) charge state of $^{229\mbox{\scriptsize m}}$Th \cite{Wense3}.
Then the ions are neutralized (either by collection on a metal surface \cite{Seiferle2} or by charge exchange in a gas or foil) and decay within microseconds via internal conversion (IC) under the emission of an electron \cite{Wense, Seiferle}.
The kinetic energy of the IC electron equals the difference of the energy of the isomer and the electron's binding energy.
Therefore, measuring the electron's kinetic energy enables to derive the isomeric transition energy.\\

Starting from a $^{233}$U source with an activity of 290 kBq, the number of extracted Th ions in the isomeric state in our experimental setup is in the range of 200 per second.
Therefore, it is envisaged to use a magnetic bottle-type spectrometer \cite{Kruit, Yamakita} that provides a large acceptance angle of nearly 4$\pi$.
In such a spectrometer the magnetic mirror effect is used to collect and collimate the electrons adiabatically. 
The kinetic energy of the electrons can be determined either via a time-of-flight measurement or via retarding fields.
As the electrons are produced in the decay of  $^{229\mbox{\scriptsize m}}$Th with a lifetime in the range of $\mu$s,  the exact start time of the electrons is not known precisely and therefore a time-of-flight measurement is not applicable.
Therefore, retarding fields must be used for an energy determination.
In this configuration all electrons are counted that exhibit a kinetic energy large enough to surpass the applied retarding fields, hence an integrated spectrum is recorded.
Within the next chapters the spectrometer is described, which will be used for a first direct energy measurement of the isomeric transition. 

\section{Design of the Magnetic Bottle Spectrometer}\label{Tech}
	\begin{figure*}[t]
\centering
\includegraphics[width=1\textwidth]{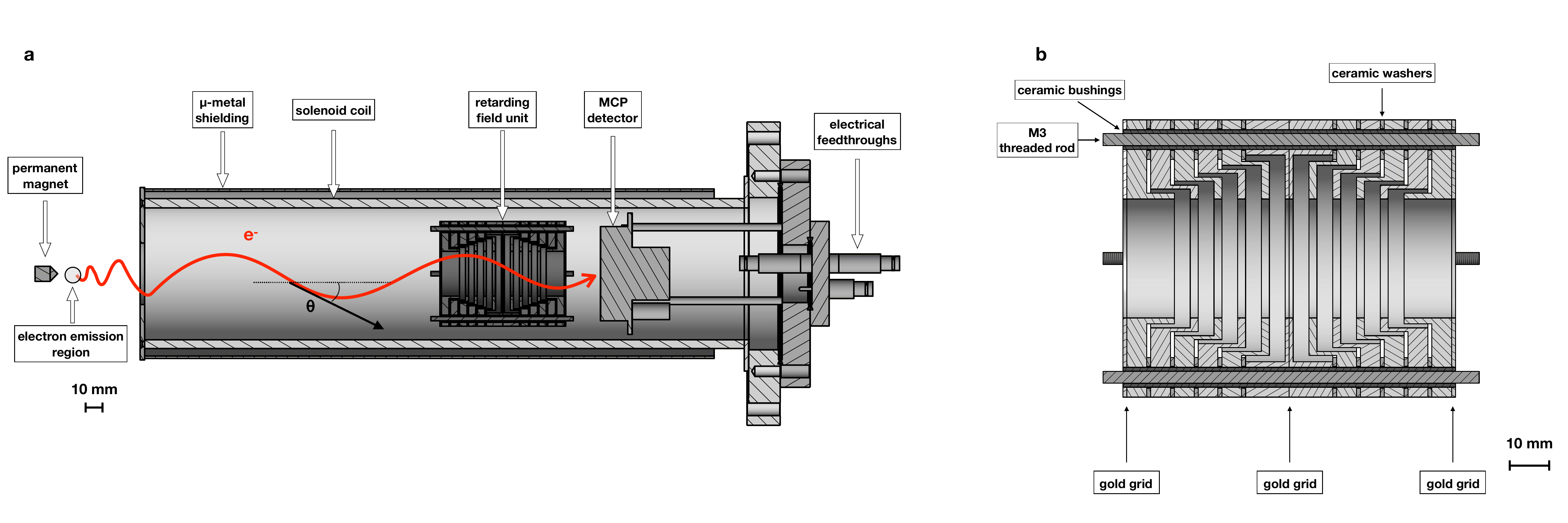}
\caption{Sectional view of the magnetic bottle spectrometer. The left panel (\textbf{a}) shows an overview of the spectrometer, with the permanent magnet, the coil, the retarding field unit and the MCP detector. The region of electron emission as well as a helical electron trajectory with its pitch angle $\theta$ is indicated. The right panel (\textbf{b}) shows a detailed view of the retarding field unit. The positions of the gold grids are indicated by arrows. \label{SectionalView}}
\end{figure*}
The design of the magnetic bottle spectrometer follows the concept that is used in Ref. \cite{Yamakita}.
In principle four components are needed to realize a magnetic bottle-type retarding field spectrometer:\\
\textit{(i)} a strong and inhomogeneous magnetic field (generated by a permanent magnet),\\
 \textit{(ii)} a flight region with a weak and homogeneous magnetic field (generated by a solenoid coil),\\
 \textit{(iii)} electrical retarding fields to block electrons, and\\
 \textit{(iv)} an electron detector.\\
A sectional view of the spectrometer setup that is studied in the following is shown in Fig. \ref{SectionalView}.
All components of the spectrometer are fixed to a DN 150 CF flange, which also carries all necessary electrical feedthroughs.
During the measurements, the spectrometer is attached to a vacuum chamber, which is pumped by a turbomolecular pump (600 l/s) and a roughing pump.
The base pressure reaches the lower $10^{-8}$ mbar region. 
All parts have been manufactured from non-magnetic materials (either aluminum or titanium) if not stated otherwise.
The region of electron emission ($\approx$2 mm above the permanent magnet) is shielded from electrical stray fields with solid aluminum. 

For electron counting a multichannel plate (MCP) detector (Hamamatsu, Type F-2223) is used.
In the next sections the generation of the magnetic and electric fields is detailed. Additionally, simulations of the retarding field unit are presented.

		\subsection{Magnetic Fields}\label{Mag}		

\begin{figure*}[tb]
\centering
\includegraphics[width=0.45\textwidth]{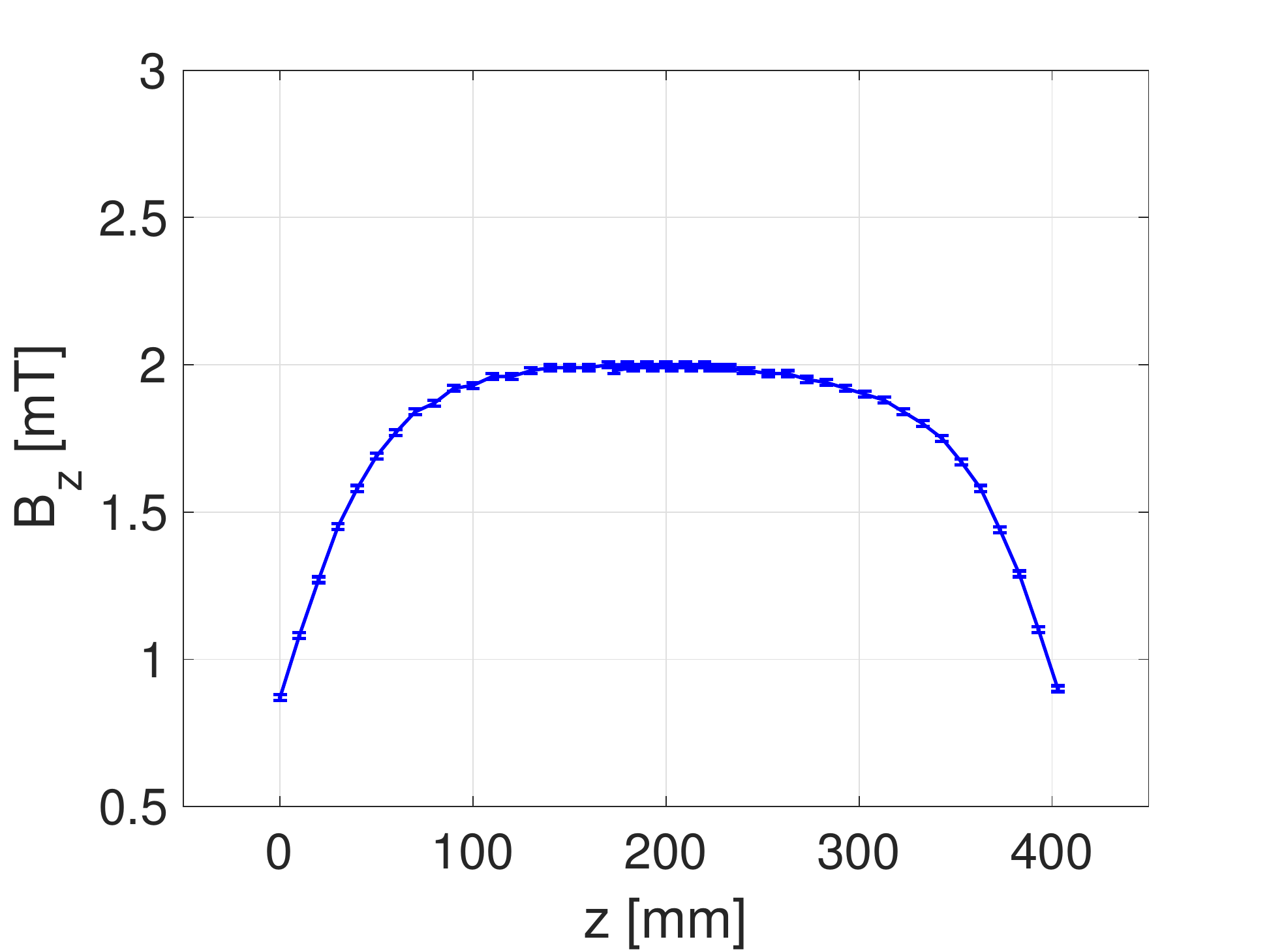}
\includegraphics[width=0.45\textwidth]{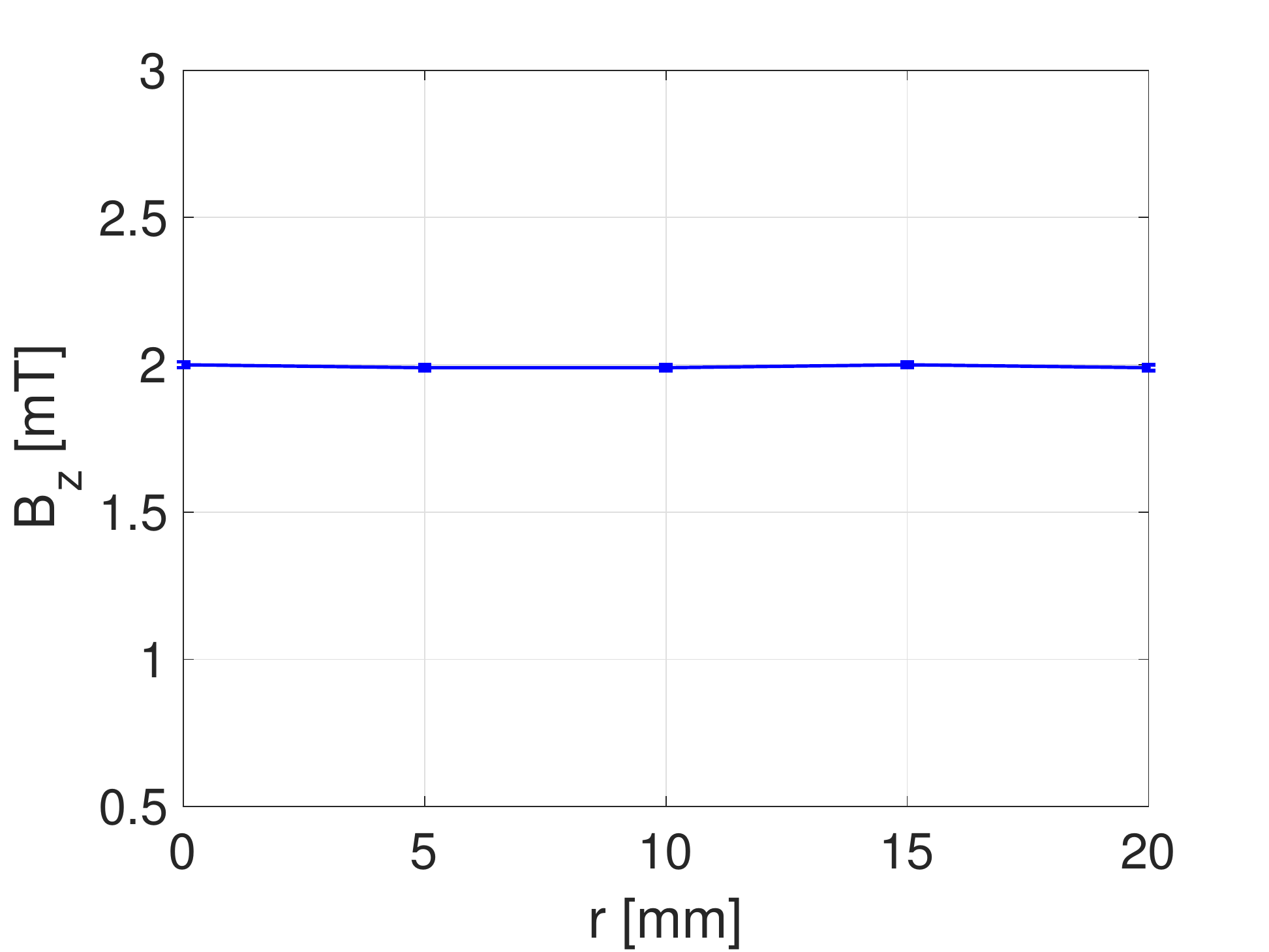}
\caption{Measurement of the magnetic field along the central axis of the coil (left) and radially in the center of the solenoid coil (right). The current supplied to the coil was 2 A. \label{MagMeasurement}}
\end{figure*}

Magnetic fields are generated by a strong permanent magnet (\textit{Vacuumschmelze Hanau, VACODYM 1.4 T}) with \O 10 mm and 10 mm height. 
In order to focus the magnetic fields to the collection region, a soft iron cone is attached on top of the magnet.
The strength of the magnetic field 2 mm above the cone tip has been measured to be  $\approx$200 mT.
The homogeneous magnetic field is generated by a Kapton isolated copper wire, which is wound around an aluminum body (length 400 mm, inner diameter 80 mm, outer diameter 90 mm, 368 windings).
The coil is placed in vacuum.
A measurement of the magnetic field inside the coil is shown in Fig. \ref{MagMeasurement}.
For a proper shielding of magnetic stray fields, the solenoid flight tube is surrounded by a 3 mm thick $\mu$-metal layer.\\
Electrons are collimated by the magnetic fields. 
The quality of the electron beam collimation depends on the ratio of the strong magnetic field in the region of electron emission and the weak magnetic field which is present in the flight tube.\\
For a quantitative assessment we assume that an electron is initially emitted under a pitch angle $\theta_i$ (see Fig. \ref{SectionalView}) above the permanent magnet in a magnetic field of strength $B_i$.
The angle $\theta$ is measured with respect to the rotational symmetry axis of the spectrometer.
After collimation, the angle $\theta_f$ can be calculated as a function of the weak magnetic field $B_f$:
\begin{equation}
\theta_f = \arcsin\Bigg[\sqrt{\frac{B_f}{B_i}}\sin{(\theta_i)}\Bigg].
\end{equation}
Electrons that are initially emitted orthogonal to the magnetic field lines ($\theta_i = 90^\circ$) represent the electrons with maximum angle in the collimated beam. 
The maximum opening angle of electrons in the solenoid coil can therefore be calculated via $\theta_{\mbox{\scriptsize max}}= \arcsin(\sqrt{B_f/B_i})$.
In our case the ratio of the magnetic fields is equal to $B_f/B_i \approx 2 \mbox{mT}/200 \mbox{mT} = 10^{-2}$, which results in a maximum angle of $\theta_f$=5.7$^{\circ}$.
	
		\subsection{Retarding Field Unit}\label{Elec}

\begin{figure}[tb]
\centering
\includegraphics[width=0.45\textwidth]{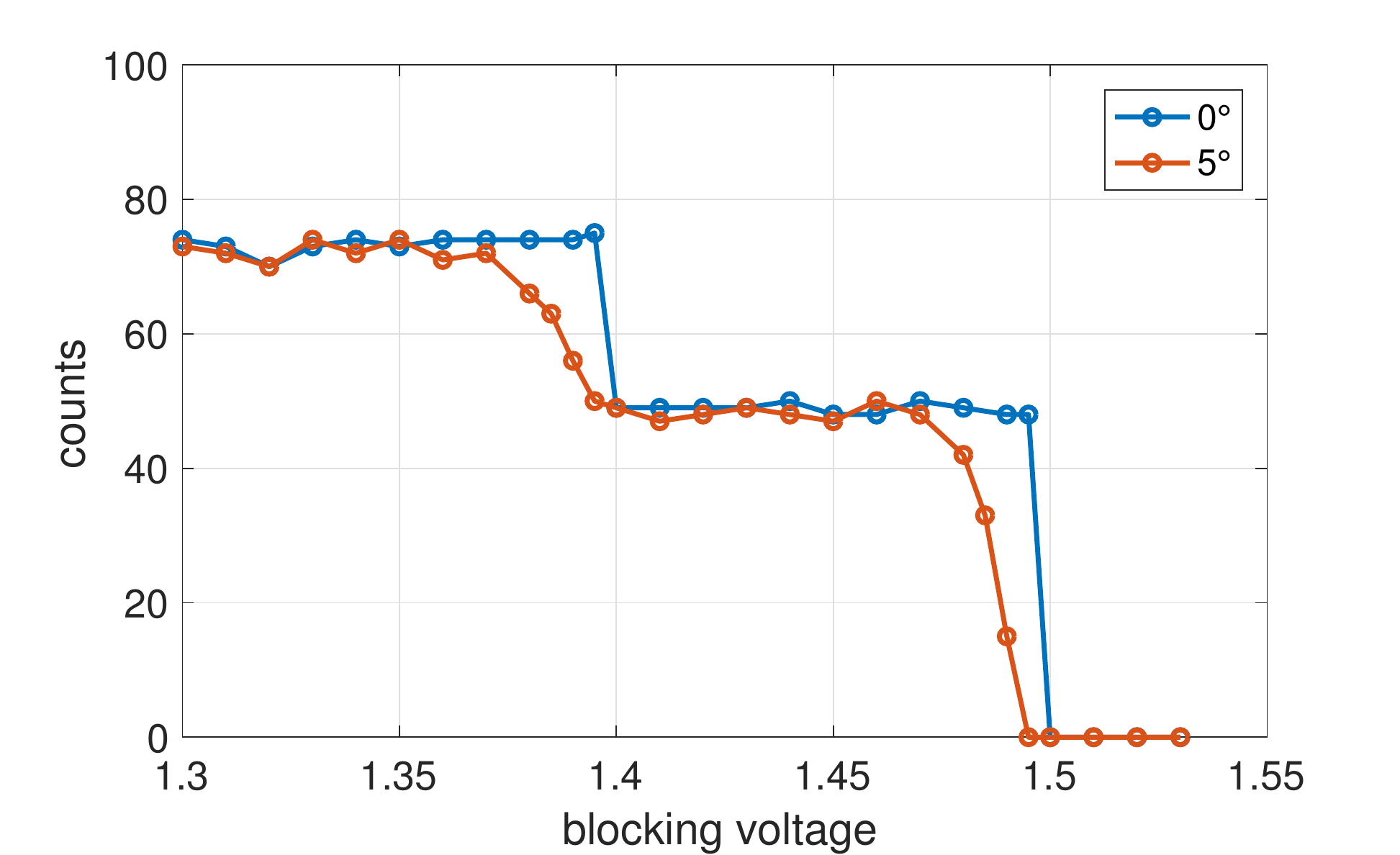}
\caption{Simion simulations of the retarding field analyzer's spectroscopic perfomance.
Electrons with different pitch angles were simulated (0$^\circ$ shown in blue, 5$^\circ$ shown in red). The retarding voltage was incremented in steps 10 mV in the plateau-like regions and 5 mV around the edges. \label{Simion}}
\end{figure}

		Electrical retarding fields are generated by a retarding field unit that consists of 12 concentric ring electrodes.
Gold grids are sandwiched between the two central electrodes and are placed at the entrance and exit of the retarding field unit.
A sectional view is shown in Fig. \ref{SectionalView}\textbf{b}.
The retarding voltage (which typically takes values between  $0$V and $-$10V) is applied to the central electrodes and the entrance and exit electrodes are set to ground.
The remaining ring electrodes are supplied via a voltage divider chain to ensure a smooth field gradient. \\
Spectral performance of the retarding field unit 
is investigated with SIMION\cite{SIMION}.
In these simulations the magnetic field of the coil is calculated via the Biot-Savart law (with the coil geometry and the current applied to the wire as input). 
The magnetic field of the permanent magnet is omitted, as it has been measured to be negligible in the region of the retarding field analyzer.
The simulations were performed with two discrete electron energies, 1.4 eV and 1.5 eV, with relative intensities of 1 and 2, respectively. 
A total number of 75 electrons are flown simultaneously through the retarding field unit and the number of electrons that surpass the retarding fields are plotted versus the (blocking) voltage applied to the central electrode.
The resulting spectrum is shown in Fig. \ref{Simion}.\\
Two scenarios are compared:
the first assumes the ideal situation of a fully collimated electron beam ($\theta_f=$0$^\circ$, blue curve in Fig. \ref{Simion}). In the second scenario only electrons with a pitch angle of $\theta_f=$5$^\circ$ are flown through the spectrometer (red curve in Fig. \ref{Simion}.
It is clearly visible that the electrode geometries do not introduce any distortion to the spectrum of the ideal electron beam. 
Electrons with a non-zero pitch angle, however, are shifted and smeared, which can be explained as follows:
As the retarding field analyzer measures the velocity component of the electrons parallel to the voltage gradient, the edges are shifted towards lower energies for the electrons with a non-zero pitch angle.
The broadening of the edge is due to inhomogeneities in the electric fields at the edges of the electrodes.\\
In the experiment, the pitch angles of the electrons are distributed between 0 and $\theta_{\mbox{\scriptsize max}}$ ($\approx$ 6$^\circ$, see section \ref{Mag}).
The broadening accounts to below 50 meV, which is below 4\% of the initial kinetic energy of the electron and can be assumed to be the limiting factor of the spectrometer's precision.
Further improvement can be achieved by placing the outer gold grids on the inside of the outer ring electrodes.
This would improve the homogeneity of the electrical fields and also the energy resolution of the spectrometer.

\section{Test Measurements}\label{Test}

In the following chapter characterization measurements of the spectrometer are presented.
In these measurements a calibrant gas is ionized by a helium gas discharge lamp (\textit{Specs, UVS 10/35}).
The calibrant gas is leaked into the vacuum chamber via a piezo valve.
No special nozzle is used and spectra are recorded typically at a (calibrant) pressure between 1$\times10^{-5}$ mbar and 4$\times10^{-5}$ mbar.

The retarding voltage is swept continuously between two voltages [$U_{\mbox{\scriptsize min}}$, $U_{\mbox{\scriptsize max}}$] in a sawtooth shape with period $t_{\mbox{\scriptsize cycle}}$.
Electrons are counted in time bins of width $\Delta t$, which are then translated to voltage bins of width $\Delta U  = (U_{\mbox{\scriptsize max}}-U_{\mbox{\scriptsize min}})\Delta t/t_{\mbox{\scriptsize cycle}}$.

The spectra are calibrated against a spectral line in the calibrant gas photo-electron spectrum and shifted in energy to correct for surface potential differences (typically ranging from to $-$0.2 to $-$0.5 eV).
The raw data can be re-binned or smoothed via a moving average.
In the plots shown in the next sections the raw data is shown as red dots, while the smoothed data is shown in solid red.\\
The measurements show the counts $N_i = N(U_i)$ obtained in the voltage bin [$U_i, U_i+\Delta U$].
The differential spectra ($D_i = D(E_i)$) are generated via
\begin{eqnarray}
D_i &=& - (N_{i+1}-N_i), \\
E_i &=& \frac{U_i+U_{i+1}}{2} = U_i + \frac{\Delta U }{2}.
\end{eqnarray}
	\subsection{Energy Resolution}\label{Res}

\begin{figure*}[tb]
\centering
\includegraphics[width=0.45\textwidth]{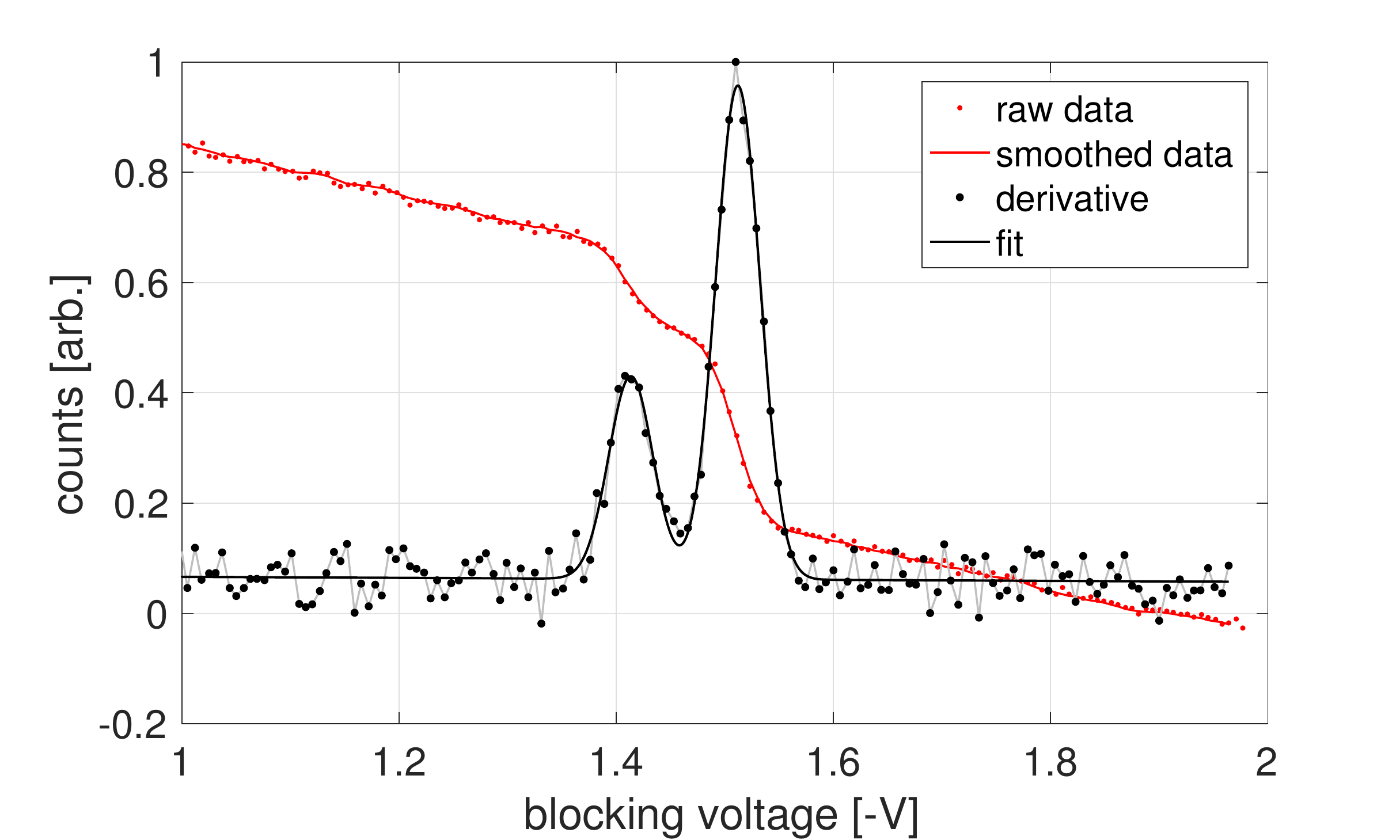}
\includegraphics[width=0.45\textwidth]{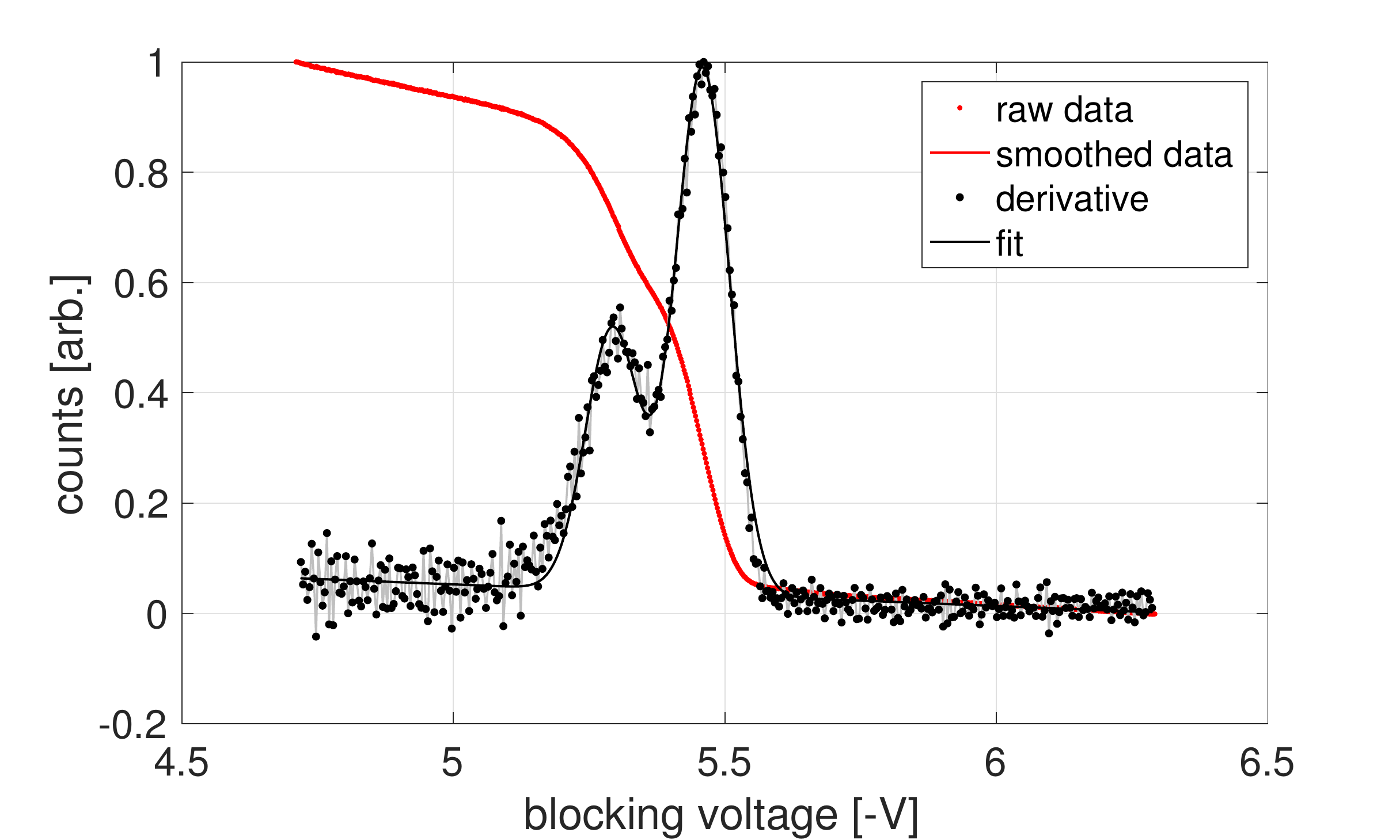}
\caption{Electron spectra recorded with neon (left) and argon (right) as calibrant gases ionized by a helium discharge lamp.   
For better visualization, the raw data counts, as well as the derivative counts, were normalized to their maximum. The voltage bin width after rebinning accounts to 6 mV and 4 mV for the neon and argon measurement, respectively. Note that for the argon measurement the difference between the raw data and the smoothed data is not visible. \label{NeMeas}}
\end{figure*}

The energy resolution of the spectrometer was measured with argon and neon as calibrant gases.\\
When argon is ionized by He I$\alpha$ radiation ($E_\gamma = $21.218 eV) it emits electrons with a kinetic energy of 5.459 eV (Ar$^+$ $^2$P$_{3/2}$) and 5.281 eV (Ar$^+$ $^2$P$_{1/2}$), respectively.
For the Ar$^+$ $^2$P$_{3/2}$  line the full width at half maximum is measured to $\approx$0.12 eV. 
The number of electrons measured in the peaks accounts to 1.9 $\pm$ 0.1 $\times 10^6$ (Ar$^+$ $^2$P$_{3/2}$) and 1.0 $\pm$ 0.2  $\times 10^6$ (Ar$^+$ $^2$P$_{1/2}$).
\\
For neon He I$\beta$ radiation ($E_\gamma = $23.085 eV) is used, which is emitted from the same UV lamp (the intensity is 2.5\% of the intensity of the He I$\alpha$ line).
Electrons with kinetic energies of 1.522 eV (Ne$^+$ $^2$P$_{3/2}$) and 1.425 eV (Ne$^+$ $^2$P$_{1/2}$) are emitted in this case.
The linewidth of the Ne$^+$ $^2$P$_{3/2}$ line at has been measured to $\approx$0.05 eV (FWHM).
The total number of electrons in the peaks is 11 $\pm$ 1  $\times 10^3$ (Ne$^+$ $^2$P$_{1/2}$) and 25 $\pm$ 1 $\times 10^3$ (Ne$^+$ $^2$P$_{3/2}$).
The measurements are shown in Fig. \ref{NeMeas}.
This leads to a relative energy resolution of 2.2\%  (3\%).\\

	\subsection{Linearity}
\begin{figure*}[tb]
\centering
\includegraphics[width=0.8\textwidth]{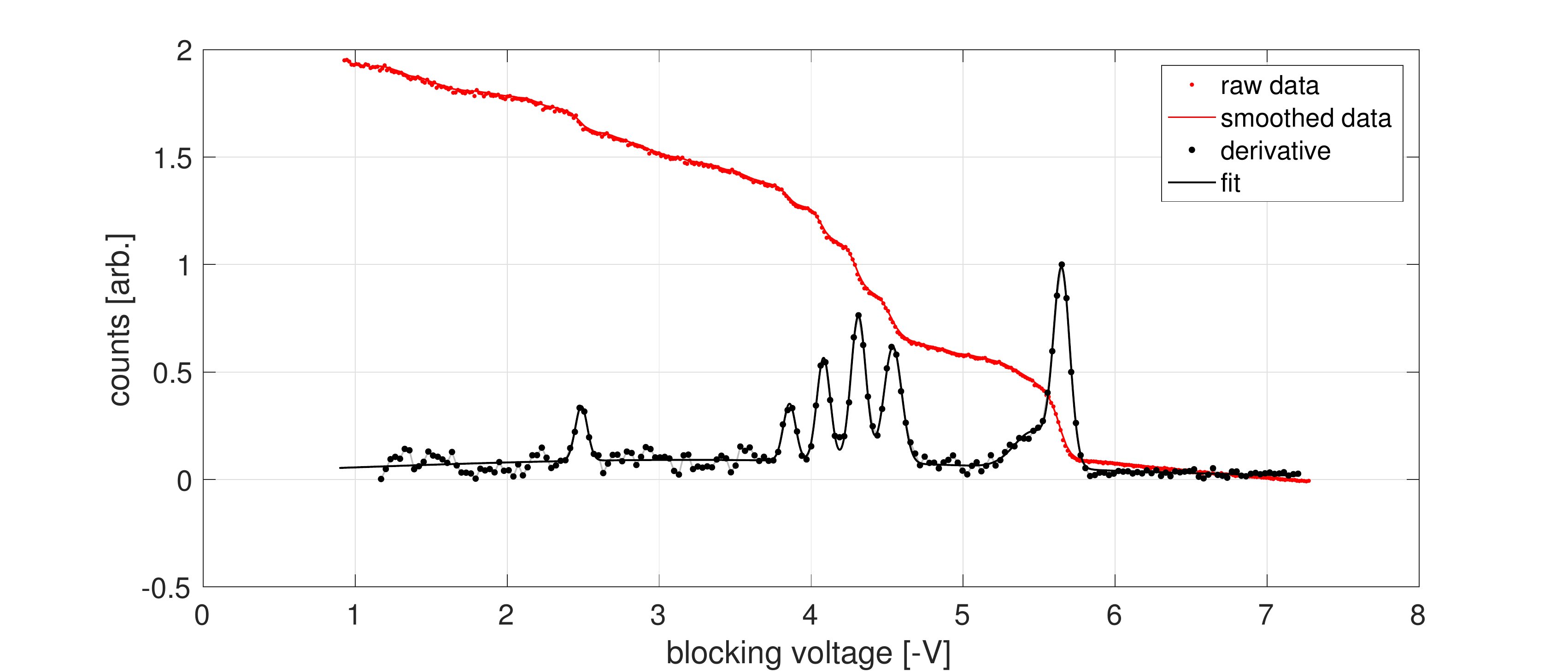}
\caption{Recorded spectrum of the vibrational molecular states of nitrogen. The raw data counts, as well as the derivative counts, were normalized to their maximum. For better visualization, the data was multiplied by two. Electrons were counted in bins of 31 mV.\label{N2Meas}}
\end{figure*}

\begin{figure}[tb]
\centering
\includegraphics[width=0.5\textwidth]{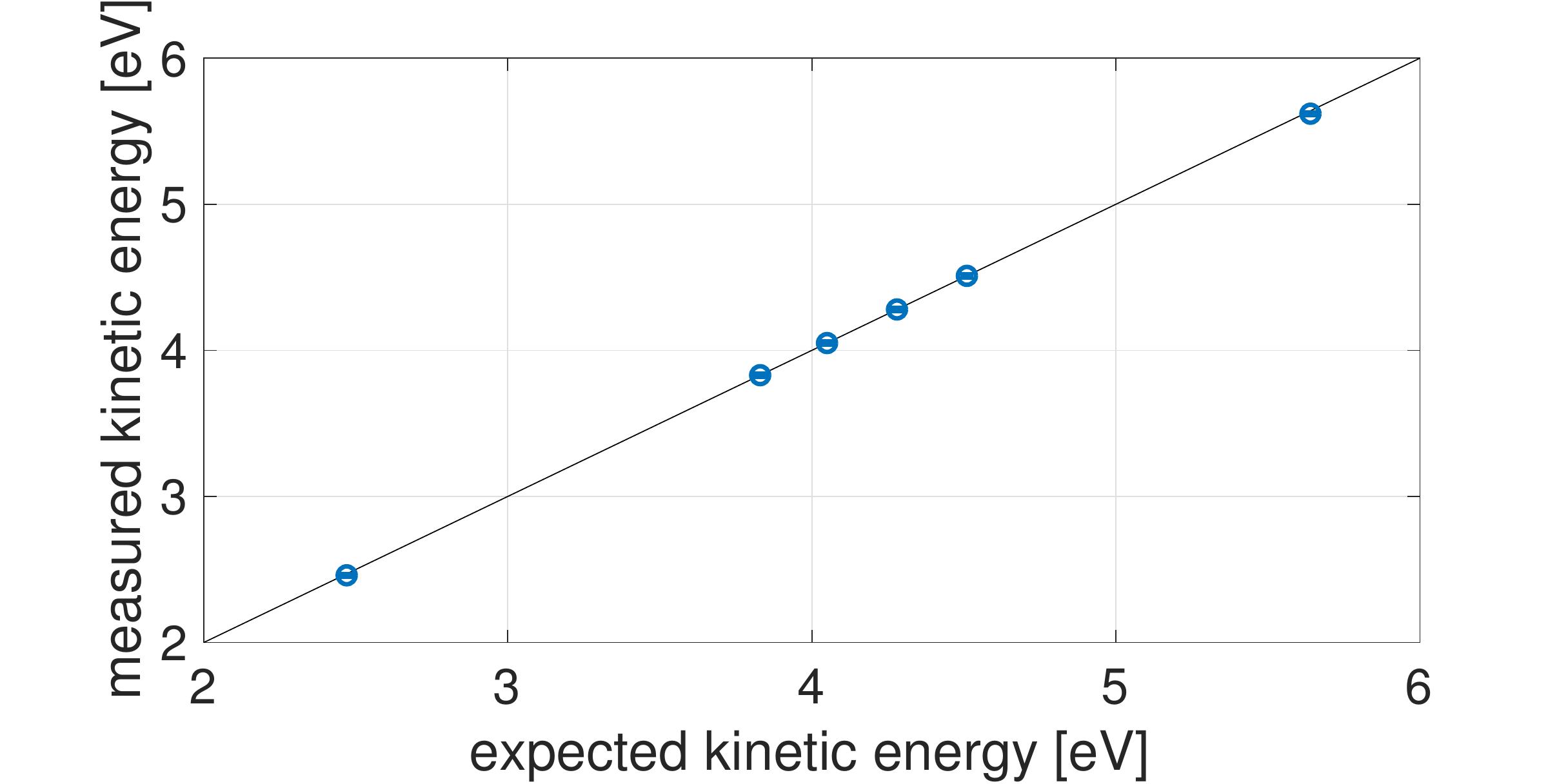}
\caption{The measured energy plotted versus the literature value. Deviations from the literature value account to below 0.5 \%.\label{Linearity}}
\end{figure}

	The spectrometer is investigated further by leaking air into the spectrometer collection region and measuring the spectrum of molecular nitrogen (N$_2$).
The spectrum was calibrated against the A$^2\Pi_u$ ($v=0$) line (E = 4.52 eV) and is shown in Fig. \ref{N2Meas}.
Vibrational states of N$_2^+$ are resolved and a linear behaviour over a range of several electron volts is measured. 
The corresponding plot, where the measured values are compared to the values measured in \cite{Gardner} is shown in Fig. \ref{Linearity}.
The vibrational states A$^2\Pi_u$ ($v=0,1,2,3$) between 3.83 eV ($v = 3$) and 4.51 eV ($v=0$) are clearly resolved. 
There is an indication for the A$^2\Pi_u$ $v=4$ state at 3.61 eV, it is, however, too weak in intensity to be clearly visible.
The spacing between the vibrational peaks is measured to be 0.23$\pm$0.02 eV. 
This value is well in agreement with the literature value of 0.229 eV (calculated with the Dunham coefficients, which can be found in \cite{Harada}).

\section{Conclusion \& Outlook}
We present a magnetic bottle type retarding field electron spectrometer for the direct measurement of internal conversion electrons emitted in the ground state decay of $^{\mbox{\scriptsize229m}}$Th.
The spectrometer has been simulated, commissioned and tested.
For the test measurements different calibrant gases were ionized with a helium discharge lamp and spectra of the electrons kinetic energies were recorded.
A relative energy resolution of better than 3\% is reached.
This will allow to determine the kinetic energy of the internal conversion electrons with an expected energy of $\approx$1.5 eV to better than 50 meV and provide a first precise direct energy measurement of the excitation energy of $^{\mbox{\scriptsize229m}}$Th.
This will be the doorway to developing a laser system able to optically control the thorium isomer, thus forming the basis for the realization of the long-sought nuclear clock.\\

\noindent This work was supported by DFG (Th956/3-2) and by the European Union's Horizon 2020 research and innovation programme under grant agreement No 664732 "nuClock".

\end{document}